\documentclass{emulateapj} 
\usepackage{latexsym}
\usepackage{mathrsfs}
\usepackage{times} 
\usepackage{graphics}
\usepackage{graphics}

\newcommand{\xim}{{\tt XIM} }
\newcommand{\apec}{{\tt APEC} }
\newcommand{\aped}{{\tt APED} }
\newcommand{\atomdb}{{\tt ATOMDB} }
\newcommand{\xspec}{{\tt XSPEC} }
\newcommand{\isis}{{\tt isis} }
\newcommand{\ciao}{{\tt ciao} }
\newcommand{\sherpa}{{\tt sherpa} }
\newcommand{\marx}{{\tt MARX} }
\newcommand{\ftools}{{\tt ftools} }
\newcommand{\flash}{{\tt FLASH} }
\newcommand{\saoimage}{{\tt ds9} }
\newcommand{\chandra}{{\em Chandra} }
\newcommand{\ixo}{{\em IXO} }
\newcommand{\xmm}{{\em XMM-Newton} }
\newcommand{\simx}{{\tt simx}}
\newcommand{\xms}{XMS}
\newcommand{\mnrasl}{MNRAS Letters}

\begin{document}
\shorttitle{XIM: An X-ray Simulator} \title{XIM: A virtual X-ray
  observatory for hydrodynamic simulations}

\shortauthors{Heinz \& Br\"{u}ggen}
\author{{Heinz}, S.}
\affil{University of Wisconsin-Madison}
\affil{Department of Astronomy, 6508 Sterling Hall, 475 N.~Charter St., Madison, WI 53593}
\email{heinzs@astro.wisc.edu}
\author{Br\"{u}ggen, M.}
\affil{Jacobs University Bremen, PO Box 750561, 28725 Bremen, Germany}

\begin{abstract}
  We present a description of the public code \xim, a virtual X-ray
  observatory.  \xim can be used to convert hydrodynamic simulations
  of astrophysical objects, such as large scale structure, galaxy
  clusters, groups, galaxies, supernova remnants, and similar extended
  objects, into virtual X-ray observations for direct comparison with
  observations and for post-processing with standard X-ray analysis
  tools.  By default, \xim simulates \chandra and the International
  X-ray Observatory \ixo, but can accommodate any user-specified
  telescope parameters and instrument responses.  Examples of \xim
  applications include virtual \chandra imaging of simulated X-ray
  cavities from AGN feedback in galaxy clusters, kinematic mapping of
  cluster velocity fields (e.g., due to mergers or AGN feedback), as
  well as detailed spectral modeling of multi-phase, multi-temperature
  spectra from space plasmas.
\end{abstract}

\section{Introduction}
\label{sec:introduction}
As the peaks of the matter distribution in the cosmic web of structure
formation, clusters hold a special place in cosmology.  The
ever-increasing power of computers has made direct hydrodynamic
simulation of complex astrophysical systems a reality.  We can now
simulate large cosmological boxes that include massive clusters of
galaxies and incorporate complex physical phenomena (such as star
formation, the energy input from black holes, metallicity injection
from supernovae, to name a few).

X-ray observations of clusters have proven to be a rich and powerful
tool to study large scale structure.  The launch of the {\em Chandra
X-ray Observatory} in particular has revealed a wealth of new
information on cluster dynamics and at the same time, posed a number
of new puzzles, such as the question of how clusters maintain their
core temperature against catastrophic cooling.

While hydro simulations can address many of these questions, it is
vital that the output data from such simulations can be compared with
X-ray observations in the most direct and faithful way possible.
Generally, galaxy cluster are simple to simulate compared to, for
example, galaxies, where complex ISM physics and star formation
seriously complicate modeling.  For that reason, it is much easier to
produce realistic simulations of clusters that can be faithfully
compared with observations. 

This comparison requires computer codes that can turn the output from
a hydro simulation into a virtual observation with a given X-ray
telescope --- a virtual X-ray observatory.

In this paper, we present such a virtual telescope, the IDL code \xim,
which is now publically available at the web site \\ {\tt
http://www.astro.wisc.edu/$\sim$heinzs/XIM}

\xim is best suited to process output from Eulerian (grid--based)
hydro codes and its direct application is the virtual observation of
thermal emission from cosmic gas.  While it was written to directly
simulate \chandra and the next generation X-ray observatory \ixo, it
can be adapted to simulate any arbitrary X-ray telescope for a set of
user-supplied response matrix files.

\begin{figure}[tbh]
\resizebox{\columnwidth}{!}{\includegraphics{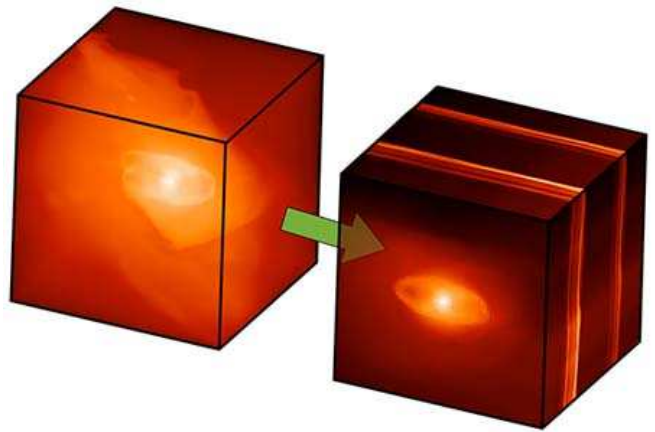}}
\caption{Sketch of \xim process: Turn a 3D data cube of density,
  temperature, velocity, metallicity, and filling factor into a (2+1)D
  spectral data cube (shown is an \ixo simulation of the iron
  K$\alpha$ line region of a radio galaxy like Cygnus A, see
  \S\ref{fig:chandra_a}).  The left cube is 450 kpc on a side, while the
  right cube has the same spatial dimensions on the x- and z-axes and
  spans an energy range from 6.2 to 6.8 keV on the y-axis (into the
  paper).\label{fig:ixosketch}}
\end{figure}

Since X-ray telescopes are imaging spectrographs (each photon is
time-- and energy--tagged, thus allowing a time-- and energy--resolved
image to be constructed), the output from a virtual X-ray observation
is a spectral imaging cube (such as would be obtained from an integral
field spectrograph in optical spectroscopy, for example).  Figure
\ref{fig:ixosketch} demonstrates the operation of \xim in calculating
such a data volume.

As the astronomical community considers the possibilities presented by
\ixo, an ultra--high resolution spectrograph with unprecedented
collecting area, it is vital that we construct detailed models of what
we can expect in terms of spectral line signatures from complex
systems like galaxy clusters and supernova remnants.  Hydrodynamic
simulations and subsequent virtual observation with \ixo present one
of the best ways to explore the incredible wealth of information such
data will offer.  

Given that no observatory with anywhere near as good a spectral
resolution, let alone combined with the planned large area nd and high
angular resolution, has ever successfully flown, experience from past
X-ray missions is clearly limiting our imagination, and more
sophisticated tools will be necessary to properly plan for the advent
of \ixo.

At the same time, long \chandra observations of diffuse objects can
benefit greatly from prior numerical simulation and it is generally
recommended that observers perform ray-tracing simulations when
proposing long observations.  \xim can perform such simulations and
presents a powerful tool in constructing mock \chandra observations of
possible targets.

This paper is organized as follows: In \S\ref{sec:implementation}, we
will describe the computational details of how \xim converts input
data into virtual X-ray observations, while \S\ref{sec:discussion}
presents some examples of how \xim could be used and a brief
discussion of some of the limitations inherent in \xim.  Section
\ref{sec:conclusions} presents a brief summary.

\section{Implementation and algorithms}
\label{sec:implementation}

In this section, we will describe the X--ray imaging pipeline {\tt
  XIM}.  \xim consistst of a suite of publically available IDL
programs that automate the creation of simulated X-ray data for a
range of satellites, with a particular focus on simulations of
\chandra and \ixo observations.

It takes input from hydro simulations of astrophysical objects with a
focus on galaxy cluster-- and cosmological simulations.  Combined with
a catalog of hydro or MHD simulations, it can thus be easily made into
a virtual observatory that can be scripted and operated remotely
(e.g., via a web interface).

In the following we describe in sequence the steps taken by \xim
towards a virtual X-ray observation.

\subsection{Data input}
\xim is directly suited for grid-based hydro simulations.  All input
must be gridded on rectilinear grids.  Adaptive mesh refinement (AMR)
is only possible in the sense of staggered meshes, not yet in the
sense of completely independent cell sizes.

The main input variables for \xim are the particle density, the gas
temperature, and the gas velocity in the form of three dimensional
arrays.  Additionally, \xim can accept arrays of metal abundance
(relative to solar) and volume filling factor of the emitting gas
(these variables can also be be passed as scalars, in which case they
will be applied to the entire simulation).

In addition, physical coordinate vectors (and optional telescope
pointing, roll angle, and off-set parameters) are used to calculate
the emitting volume and angular size of each volume element of the
input data, given an object red-shift and cosmology.  By default, \xim
uses concordance cosmological parameters of $H_{0}=70\,{\rm km/s}$,
$\Omega=1.0$, $\Omega_{\Lambda}=0.7$, $\Omega_{\rm matter}=0.3$.

The simulation output is calculated on a user specified energy grid,
which can be set to correspond one--to--one to the energy channel
mapping of the instrument.  This allows the user to easily create full
resolution spectra as well as multi-color images, or low-resolution
spectra for reduced computational expense (high resolution spectra at
\ixo resolution can be very costly to run).  A uniform turbulent
velocity broadening can be applied to the spectrum to simulate
sub-grid effects.  A uniform foreground hydrogen column density can be
specified for photo-electric absorption.


\subsection{Spectral models}
\label{sec:spectra}

\subsubsection{Thermal spectral calculation}
By default, \xim uses the publically available \apec model to
calculate spectra from thermal plasma emission (assuming gas in
coronal equilibrium) for each cell \citep{smith:01}.  \apec
self--consistently calculates the equilibrium ionization balance for a
thermal plasma.  Atomic data are taken from the \atomdb using \aped
\citep{smith:01b} and combined with bremsstrahlung continuum for all
species.  \apec output is provided as a table of spectral models.

Given the fact that \apec itself provides a table of model spectra,
and given that the computational expense of extracting a spectrum from
\apec is high, the most economical method of calculating a large
number of spectra with \xim is by creating an oversampled table of
model spectra for the entire temperature range spanned by the
simulation.  Thus, the spectral projection in \xim is based on
interpolation on a table of model spectra, on a logarithmic
temperature and photon energy grid (see \S\ref{sec:projection}).

The default temperature binning of the table is 66 bins per decade in
temperature (significantly oversampled with respect to the \apec model
output and sufficient for high-accuracy spectral interpolation, again
see \S\ref{sec:projection}).

\begin{figure}
\resizebox{\columnwidth}{!}{\includegraphics{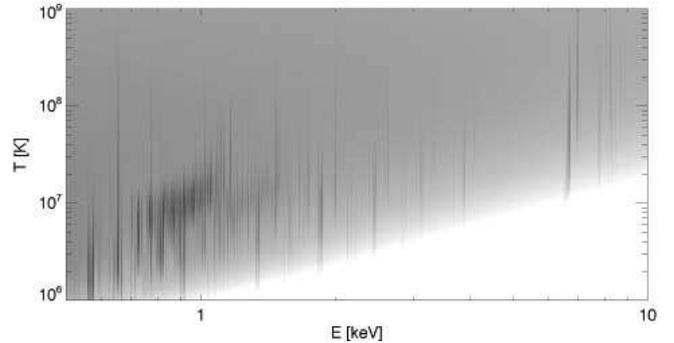}}
\caption{Twodimensional image of a model grid at default \xim
  resolution for \chandra simulations, as function of gas temperature
  and photon energy, for solar abundance, as calculated from \apec.\label{fig:apec}}
\end{figure}

Two sets of spectral tables are created: One for Helium and Hydrogen,
assuming primordial cosmic Helium abundance, and one for heavier
elements (with relative abundance of metals fixed to the solar
ratios).  The metal abundance $Z$ can be specified for each cell.  A
model grid at default \chandra resolution is shown in
Fig.~\ref{fig:apec}.

\subsubsection{Other spectral models} 
\xim is not limited to the implementation of \apec described above for
spectral modeling.  Any user-defined spectral model can be applied
(e.g., simple powerlaw spectra) and an arbitrary set of spectral
parameters and input data can be passed.  

However, because \xim is optimized for thermal emission, the spectral
model must be such that the emission is proportional to some power of
a density variable (this could be relativistic electron pressure,
which enters synchrotron emission as the 7/4 power) and a temperature
variable (which controls the spectral shape and normalization, e.g.,
synchrotron age).

The output spectra must be provided in the form of a two-dimensional
table in temperature and photon energy with logarithmic spacing on the
energy axis

\subsubsection{Foreground absorption}
\xim allows the specification of the neutral Galactic hydrogen column
density to calculate photo-electric foreground absorption following
the Wisconsin Absorption Model \citep[WABS][]{morrison:83}.


\subsection{Spectral projection}
\label{sec:projection}

The spectral projection along the line of sight is performed along an
arbitrary viewing angle.  For each cell of the input data, a unique
spectrum is calculated for the cell temperature, cell density, and its
proper Doppler shift, given the line--of--sight velocity of that cell.

To project along an arbitrary line--of--sight, the input data arrays
are rotated (using tri-linear interpolation) onto a new grid that
orients the x-axis along the line--of--sight vector.  The input
velocity arrays are also projected onto the line--of--sight vector to
calculate a line--of--sight velocity for every cell.  Line--of-sight
integration then simply proceeds along the x-axis of the new array.

\subsubsection{Doppler shift and energy gridding}
The Doppler frequency correction from the emitting to the observed
frame is
\begin{equation}
  \nu_{\rm obs}=\delta \nu_{\rm emit}={\nu_{\rm emit}}\sqrt{\frac{1 -
      v_{\rm LOS}/c}{1 + v_{\rm LOS}/c}}
\end{equation}
If the frequency grid is logarithmically spaced, Doppler shift is a
simple addition.  This suggests that spectral calculations should be
performed on logarithmic grids, which is the model adopted by \xim.
On a logarithmic energy grid, Doppler shift is a simple linear shift
along the grid, which is computationally trivial.

Thus, given an output energy grid (either user--defined or fixed to
the instrument channel map), \xim creates an over--sampled
logarithmically spaced energy grid that is padded on both sides by the
maximum Doppler shift encountered in the input data and takes into
account the energy resolution of the instrument, additionally
extending the grid to enclose a minimum of 99\% of the energy at the
grid edge.

The energy resolution of the logarithmic grid is set to oversample the
target output grid by at least a factor of 2.  Since the energy grid
used by \xim is logarithmic, most of the energy range is oversampled
by a larger factor.

Doppler shift is performed by linear interpolation between the nearest
two integer shifts along the energy axis that straddle the actual
Doppler shift value of each cell.  This accurately reproduces the mean
line centroid energy and leads to a small error in line width that is
well below the energy bin width and well below the intstrument
resolution.

\begin{figure}
\resizebox{\columnwidth}{!}{\includegraphics{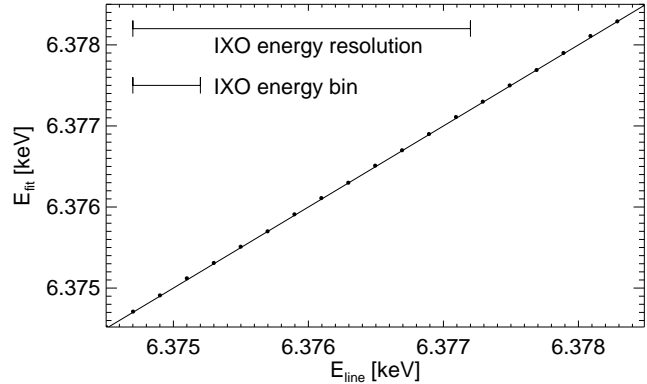}}
\caption{Accuracy of the numerical Doppler shift procedure relative to
  the actual Doppler shift.  Shown are \xspec fits to the line
  centroid (dots) for a range of Doppler shifts applied to the
  gaussian line and then virtually observed with \ixo (actual Doppler
  shifted centroid energy shown as solid line).  The centroid is
  determined to very high accuracy.  Also shown are the \ixo energy
  resolution of 2.5 eV and energy channel width of 0.5
  eV.\label{fig:eplot}}
\end{figure}

The accuracy of this procedure is demonstrated in
Fig.~\ref{fig:eplot}, which shows the line centroid energy determined
by \xspec for a gaussian spectral line with varying Doppler shift
virtually observed with \ixo (to sufficiently high signal--to--noise
to eliminate statistical error).

\subsubsection{Line--of--sight integration}
In addition to the Doppler shift correction, the line--of--sight
integration must interpolate the spectral model for each cell from the
table provided by the spectral model calculation described in
\S\ref{sec:spectra}.

This is done by linear interpolation between the two nearest
temperature spectra in the model table.  Given the default dense
sampling of the spectral model grid (66 temperature bins per
temperature decade), this method provides a smoothly varying, accurate
description of the spectra as a function of cell temperature.

Given both temperature and Doppler calculations, the line--of--sight
integration is then performed on a cell--by--cell basis.  For an input
grid of size $N_{x}\times N_{y} \times N_{z}$ and a logarithmic energy
grid with $N_{\rm le}$ bins, this involves $N_{x}N_{y}N_{z}N_{\rm le}$
calculations, which can be very time consuming and lends itself to
trivial parallelization.  A simple implementation of parallelization
of this step is included in \xim.

The line--of--sight integration is performed {\em before} re-gridding
to the instrument pixel scale to preserve the maximum amount of
velocity detail from the original hydro simulation.  However, for
high-resolution spectra, as encountered with \ixo, memory requirements
can become extreme. In some cases, it is therefore advisable to
re-sample the input grid to lower resolution, appropriate for the
lower angular resolution of the telescope (however, this procedure
will lead to an under-estimate of the Doppler width of the lines).

To keep memory requirements at a minimum, spectra are accumulated on a
pixel--by--pixel basis (i.e., the line--of--sight integration is
performed along the line of sight first, then across the image).
Thus, only the input data and the spectral cube of dimension
$N_{x}N_{y}N_{\rm le}$ have to reside in memory.

\subsection{Virtual observation}
\label{sec:observation}

The output from the line--of--sight projection encapsulates the
object's cosmological red-shift, but it is still instrument--indepent
(though often the energy grid will be chosen to be suitable to a
particular instrument to keep computational expense and memory
requirements at a reasonable level; e.g., for virtual \chandra
observations, emphasis should be placed on spatial resolution, while
for virtual \ixo obsevations, the energy grid should be at the highest
resolution possible).

Starting from the output file of the line--of--sight inegration,
virtual observations with different target telescopes and instruments
are performed in a number of steps described below.

\subsubsection{PSF simulation}
\label{sec:psf}
The line--of--sight integrated spectral cube is convolved with the
telescope point spread function (PSF).  \xim allows user specified PSF
models.  Note that off-axis effects on the PSF and vignetting are
currently not included in \xim, i.e., it is assumed that the PSF is
uniform across the field--of--view.

The default PSF used for \ixo simulations was extracted from the \ixo
simulator \simx, which provides a monochromatic PSF (specified at
6keV).  

The default PSF for \chandra simulations was extracted from \marx
ray--tracing simulations and is interpolated in energy with a native
spacing of 0.5 keV.  The low-energy resolution is 0.5 arcseconds (see
\chandra users guide for more information on the energy dependence of
the PSF).  More accurate \chandra imaging simulations, including
vignetting and off-axis effects, can be achieved using the optional
\marx interface described in \S\ref{sec:chandra}.

\subsubsection{Convolution with instrument responses}
\label{sec:response}
A critical component of realistic X-ray simulations is the
incorporation of the effective area of the telescope as a function of
photon energy (encapsulating mirror effective area and detector
quantum efficiency, typically encoded in the ancillary response file
in X-ray astronomy) and the redistribution function of photon energy
(reflecting the spectral resolution of the telescope and detector,
typcially encoded in the response matrix file). This happens in three
steps:
\begin{itemize}
\item{Given a response file (following the CAL/GEN/92-002 FITS
    specifications), \xim either re-grids the telescope response to
    the logarithmic energy bin on the data energy axis and the
    user-supplied output energy grid on the the detector axis, or, if
    \xim uses native telescope energy channels, the line--of--sight
    projection is re-gridded from logarithmic onto detector channels.}
\item{Next, the line--of-sight projected, PSF--convolved data cube is
    re-gridded onto physical detector pixels for the specified
    telescope/detector.}
\item{Finally, \xim convolves the spectral cube with the instrument
    response onto the specified output energy grid, resulting in a
    spectral cube of count rates per energy bin per pixel as detected
    by the instrument.}
\end{itemize}

\subsubsection{Virtual detection}
\label{sec:detection}
\xim assumes Poisson noise for counting statistics and produces an IDL
data file that includes the detector count rate (without statistical
error) and a counts file for the user-specified exposure time that
includes Poisson errors.

In addition to IDL output, \xim can produce a FITS events file that
can be analyzed with standard X-ray astronomy tools like \ftools and
\saoimage.

\xim can also produce a single FITS spectrum for the entire
observation.  Post-processing with the spectral generator of \xim
allows the application of IDL mask arrays to select sub-regions of the
spectrum (or select different weighting for different pixels for the
combined spectrum) and produce FITS spectra that can be analyzed with
\xspec, \isis, and \sherpa.  It is possible to create multiple
different spectral files for different regions from the same
ximulation (e.g., annuli around a cluster center for deprojection
analysis).

\subsubsection{Backgrounds}
\label{sec:backgrounds}
\xim adds sky background and instrumental background to the virtual
observation. These can be user-specified spectra or standard
background estimates included for \chandra and the \ixo calorimeter
\xms.

Sky background is treated like source photons and is folded through
the instrument response (though assumed to be uniform across the
field--of--view, so no PSF convolution is performed).  Instrument
background is added directly to the output counts without convolution
with the response.

The default background for \chandra was extracted from the black sky
background file for post-2000 data.  The \ixo background estimate is
based on the most recent estimates available on the telescope wiki
page and will be continually updated as more sophisticated estimates
become available.

\subsubsection{Virtual Chandra observations with MARX}
\label{sec:chandra}
\xim comes with default \chandra observing parameters (the latest
aimpoint response files should be downloaded by the user from the
Chandra web archive).  This alows high-fidelity simulations of objects
near the aimpoint.

However, for observations of objects far off-axis, PSF effects and
vignetting should be taken into account.  Thus, for the most realistic
virtual \chandra observations, \xim uses \marx (the official \chandra
simulation software) to perform ray-tracing simulations of the
observation.  The output is again a fully compliant \chandra FITS
events file that can be analyzed using \ciao and the \chandra pipeline
like regular \chandra data.

\section{Discussion}
\label{sec:discussion}
Having briefly described the steps taken by \xim to convert a grid of
data taken from a hydro simulation into a virtual X-ray observation,
we will present some examples of possible applications (the details of
which will be presented in a separater paper) and discuss some of the
limitations.

\subsection{Examples of applications}
\label{sec:applications}
\subsubsection{\chandra imaging of radio galaxies in clusters}
\label{sec:chandra_a}
Before going into more detail, we list several examples of possible
applications of \xim for \chandra:
\begin{itemize}
\item{A good demonstration of the power of X-ray imaging is the
    unsharp--masked image of the Perseus cluster \citep{fabian:03}
    that shows a series of concentric ripples (identified as sound
    waves and a weak shock).  While direct one--to--one simulations of
    a Perseus are not possible given the difficulty in reproducing the
    details in cluster weather that likely lead to the asymmetric
    appearance of the cluster, it is nonetheless very usefull to
    examine hydro simulations for whether they can reproduce the
    observed surface brightness fluctuations induced by these waves.}
\item{X-ray astronomy makes use of object colors for selection
    effects. For example, \citep{forman:07} used a color cut on a long
    \chandra observation of the Virgo cluster to detect a weak shock
    driven into the cluster by the radio galaxy M87 at the center.}
\item{As a further example, a cosmological simulation containing
    multiple galaxy clusters can be examined for the optical set of
    bands within which to select galaxy clusters, depending on the
    instrument used.}
\end{itemize}

Multi-color imaging is particularly easy and fast using \xim, given
that only a small number of spectral channels have to be calculated
(\apec calculates the integrated photon luminosity for a given energy
band, so no interpolation is involved when calculating broad band
images).

\begin{figure}
\begin{center}
 \resizebox{\columnwidth}{!}{\includegraphics{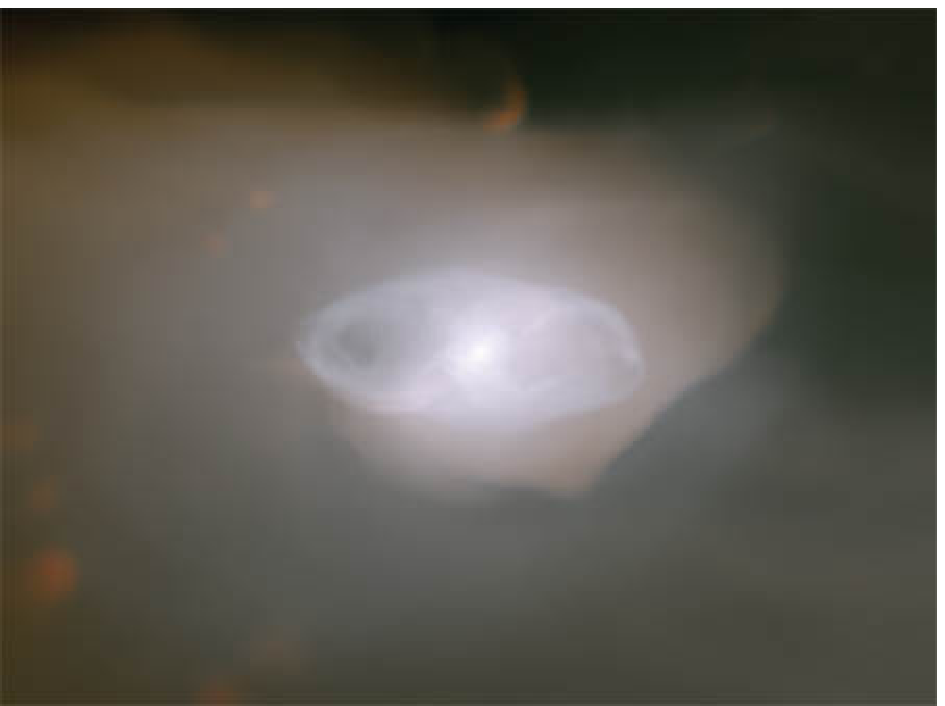}}
\resizebox{\columnwidth}{!}{\includegraphics{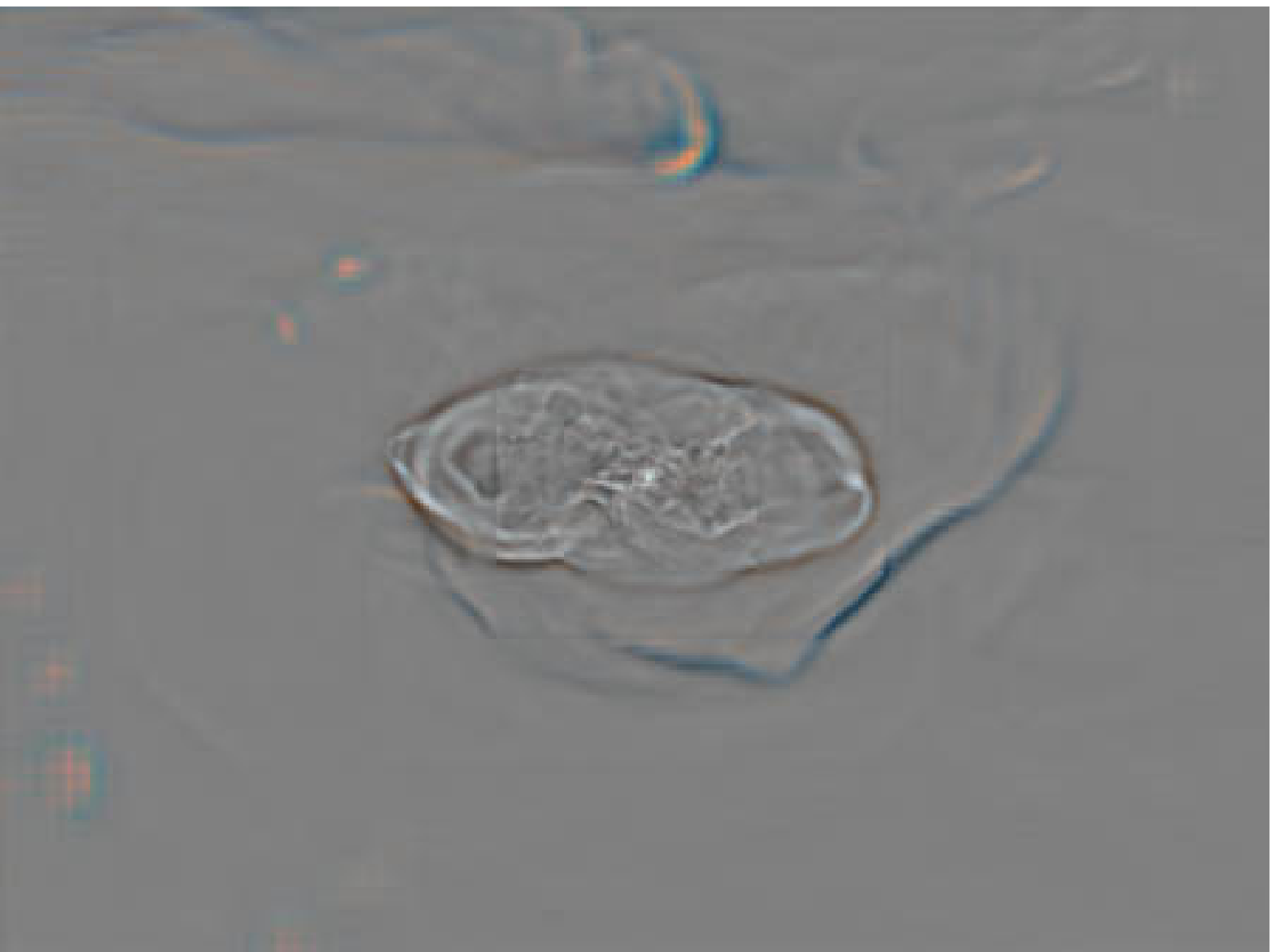}}
\end{center}
\caption{Simulated \chandra image of Cygnus A ({\em top}) and
  unsharp-masked version of the same image ({\em bottom}).  Three
  colors correspond to 0.5-2.0 keV (red), 2.0-5.0 keV (green), and
  5.0-10.0 keV (blue).  The image x and y dimensions are 440 x 330
  kpc.  The unsharp-masked image clearly shows ripples reminiscent of
  those seen in, for example, Perseus A.  Note that the images shown
  are calculated with infinite signal--to--noise to demonstrate the
  appearance of ripples in X-ray images behind shock waves.  Given a
  proper exposure time, it would be trivial to determine the
  detectability of features at a given surface
  brightness.\label{fig:chandra_a}}
\end{figure}

Fig.~\ref{fig:chandra_a} shows the simulated \chandra surface brightness
and the unsharp-masked image for a hydro simulation of Cygnus A.  The
details of the hydrodynamical simulation are described in
\cite{heinz:06b} and we limit this discussion to a very brief summary
of the features of this simulation: We used the publically available
\flash code \citep{fryxell:00} to simulate the injection of a
super-sonic, powerful jet into a galaxy cluster (taken from
\citep{springel:01}).  The jet power ($10^{46}\,{\rm ergs\,s^{-1}}$)
and the cluster temperature ($\sim 5\,{\rm keV}$) and central density
match the canonical parameters for Cygnus A very well (e.g.,
citep{wilson:02}).  When placing the simulated cluster at the
red-shift $z=0.0561$ of Cygnus A (picking a time step that reproduces
the observed angular sizes of the X-ray cavities and the shock in
Cygnus A) the simulated \chandra surface brightness is in good
agreement with the actual \chandra observation (to within about 10\%).

The power and importance of virtual X--ray imaging can be demonstrated
from the unsharp--masked image: The appearance of ripples in the wake
of the outgoing shock wave is clear from this image.  It implies that
X-ray surface brightness ripples can be excited in the wake of one
single outgoing shock wave as a result of jet dynamics and do not
necessarily require multiple episodes of activity of the central radio
galaxies.

Given the abundant data on X--ray cavities in clusters in the \chandra
archive, a statistical treatment of bubble properties has now become
possible.  When computing the observability and imaging properties of
a sample of bubble either from simulations or analytic models, it is
critical to employ faithful rendering of the virtual X-ray data.
Using \xim, \citet{bruggen:09} have demonstrated the importance of
line--of--sight projection for estimating bubble position and size
from observations.

\subsubsection{High-resolution \ixo spectroscopy of a virtual clone of
  Cygnus A} 
\label{sec:ixo_a}
Using the same hydro simulation described in \S\ref{sec:chandra_a}, we
used \xim to simulate a high--resolution spectrum of a powerful radio
galaxy like Cygnus A observed with \ixo.  A more detailed description
of the power of high--resolution spectroscopy for the study of radio
galaxies will be published in a companion paper.  Here we wish to
demonstrate the power of such observations using a simple example.

\begin{figure}
\resizebox{\columnwidth}{!}{\includegraphics{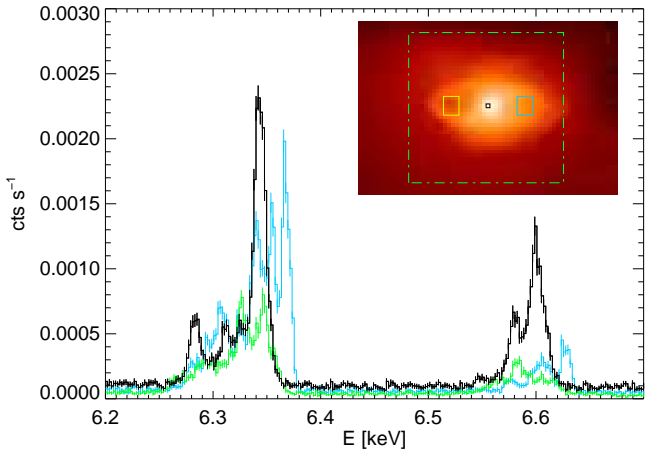}}
\caption{Simulated \ixo spectra of Cygnus A for three lines of sight
  ({\em green/yellow}: through the left cavity; {\rm black}: Through
  the cluster center; {\em cyan}: Through the right cavity).  The
  three most prominent peaks of the cyan spectrum show the approaching
  and receding cavity walls, as well as the cluster fore- and
  background emission.  The offset relative to the cluster emission
  (black) is due to some rotation in the cluster and the jet
  inclination relative to the line of sight.  The image inset shows
  the simulated image of the cluster.\label{fig:ixo_a}}
\end{figure}

The high spectral resolution of \ixo will allow detailed kinematic
mapping of the expanding cavities of radio galaxies.  Figure
\ref{fig:ixo_a} shows three spectra, one through the cluster center
and two through the expanding cavities.  The line emission is due to
the K-$\alpha$ lines from FeXXV and FeXXVI (Helium-- and
Hydrogen--like iron), which emit abundantly in hot cluster atmospheres
like that of Cygnus A.

The two spectra taken through the cavity centers show clear evidence
for Doppler-shifted line emission from the front and back walls of the
cavity and it is possible to simply read off the expansion velocity
from the spectrum to be $614\,{\rm km\,s^{-1}}$ from the separation of
the three right-most peaks of the cyan spectrum.  These three peaks
are all due to the resonance line of FeXXV from the rear wall (left
peak at $6.341\,{\rm keV}$), the front wall (right peak at
$6.367\,{\rm keV}$) and the fore-- and background cluster emission
(middle peak at $6.354\,{\rm keV}$).  The expansion line--of--sight
velocity in the simulation is $v_{\rm shell}\approx 650\,{\rm
km\,s^{-1}}$.

More sophisticated model fitting would allow us to determine the
emission measure distribution at different velocities, possibly
allowing a deconvolution of the internal dynamics of the shell (which
is not moving at uniform velocity).  This could be used to, for
example, determine the rate at which gas is flowing back along the
shell as well as detecting the creating of high entropy gas around
shells as direct evidence of heating and feedback.

Other obvious examples of using high-resolution X-ray spectroscopy in
the context of galaxy clusters include detailed \apec model fits to
determine the temperature and metallicity distribution of the cluster
gas.  Demonstrations of the power of this method are beyond the scope
of this paper, but the \xmm obervations of galaxy clusters that gave
rise to the second cooling flow problem (ultimately leading to the
current paradigm of AGN heating of clusters) provide evidence of the
power of this method \citep{peterson:03} even at the lower resolution
and throughput of \xmm.

It is also possible to incorporate and test multi-phase models versus
observations.  This can be easily achieved by super-imposing multiple
\xim simulations (one for each phase).

\subsection{Limitations}
\label{sec:limitations}
Finally, we will briefly discuss some of the current limitations on
the software that are not likely to be remedied or implemented in the
near future.

\subsubsection{Grid geometries}

Based on the projection method, the code can only handle input data on
regular grids that can be described by a single, rectungular three
dimensional matrix.  This excludes true AMR grids (though it allows
for staggered grids) and, naturally, Lagrangian formulations from
direct application.  To use \xim with AMR or SPH output data, they
will have to be re-gridded onto a regular grid.  The same is true for
non-cartesian grids.  The examples shown above were all calculated
with the AMR code \flash and then re-gridded onto a regular cartesian
grid.

\subsubsection{Line--of--sight projections}

Direct integration along the coordinate axes is computationally
simpler and faster, which is the method employed by \xim.  This
implies that all input arrays have to be rotated prior to the spectral
projection.  As with regular projections through a rectangular volume
for arbitrary line--of--sight vectors, missing values lead to the
appearance of projection effects (e.g., the volume is rendered as a
projected box.  Currently, \xim does not allow for period boxes to be
rendered without data loss.  It is possible to clip the data, however,
to avoid the appearance of projection artifacts from missing values.

\subsubsection{Spectral models}
As already described, \xim incorporates the \apec thermal plasma
emission code for plasma in coronal equilibrium.  The method used to
interpolate spectra assumes that the spectral shape depends only on
one parameter (the temperature or an equivalent variable that can be
passed instead of temperature) and that the emissivity is proportional
to the square of the density (or to some power of an equivalent
variable that can be passed instead)

\xim allows user-defined spectral models as long as they can be
written in such a way that they depend on two parameters (one that
determines the spectral shape, such as temperature and one that
affects the normalization through a powerlaw dependences). As long as
this is the case, it is straight forward to incorporate other
models. For example, synchrotron emission could be included by having
the spectral shape (index and/or cutoff) determined by one parameter
(which is passed as the variable t, but could in fact be the spectral
age of the plasma) and the overall normalization is determined by the
non-thermal pressure (which would be passed as the parameter n).
Spectral shifts would be incorporated through the line--of--sight
velocity v.

Clearly, this method is somewhat limiting in the way that spectral
models can be formulated.  For example, photo-ionized plasmas will be
difficult to implement.

\subsubsection{Imaging restrictions: Vignetting and off-axis}
As already mentioned above, \xim assumes a uniform PSF across the
field--of--view and neglects vignetting.  For \ixo, these extent of
these effects is currently not well known.  For \chandra, the best way
to incorporate these effects is to use the \marx option to pipe the
spectral cube through the telescope ray tracing software.  This will
produce the most realistic \chandra simulations.

\section{Conclusions}
\label{sec:conclusions}
We have presented a description of the public code \xim.  The code
produces virtual X-ray observation from hydro simulation of
astrophysical plasmas.  The code is best suited for the simulation of
thermal emission, using the \apec model.  \xim incorporates detailed
telescope parameters and responses for \chandra and \ixo.  Examples of
the application for \xim include multi-color imaging, detailed
surface--brightness mapping (e.g., using unsharp--masking),
high-resolution kinematic studies of the dynamics of galaxy clusters,
groups, and supernova remnants, and multi-temperature, multi-phase
high resolution spectroscopy of space plasmas. Current limitations
include the restriction to regular, rectangular grids, spectral models
that can be parameterized in a fashion equivalent to thermal emission,
and restriction to on-axis imaging performance.

\end{document}